# Organised crime's infiltration in the legitimate private economy: An empirical network analysis approach

It is estimated that Italian Mafias registered €135 billion in profits only in 2010 (SOS Impresa, 2010). Part of this huge amount of money – coming mostly from the drugs, prostitution and arms illicit markets – is often used to invest into the legitimate private economy. At the end of 2011, for instance, several former members of the Camorra – an organised crime group[1] based in Naples, Italy – confessed that the great majority of firms dealing with major economic transactions in their territory of influence is either directly or indirectly connected to the Mafioso group (Iurillo, 2011). As a consequence, the affected economy destabilises, becomes entrenched with violent forms of competition and is bound to stagnation. Such phenomena are observable also in the Italian regions of Apulia and Basilicata, where Mafioso activity represents an economic shock that diminishes the expected GDP per capita by 16% on average, according to a recent econometric study by the Bank of Italy (Pinotti, 2012).

Organised crime's infiltration in the economy is not an Italian oddity, but a global security issue. The Indian D-Company possesses a trading company in Dubai and several construction companies in the Gulf (Ray and Tare, 2011), while the Japanese Yakuza heavily invests in Asian stock markets (Grennan, 2000: 239). Nonetheless, few are the attempts to uncover the patterns followed by criminal organisations in their business ventures. Apart from a pilot study performed by the US National Institute of Law Enforcement and Criminal Study (1970), literature that empirically evaluates theories and models of organised crime's penetration in the economy (Caterini, 2011; Fiorentini and Peltzmann, 1997; and Kumar and Skaperdas, 2008 are notable examples) is nearly absent. The reason lays mostly in the poor public availability of data on criminal activity, or in the highly risky task of gather it (Cressey, 1967; Chambliss, 1975).

This paper attempts to partially fill this gap thanks to access to information about the Sicilian Mafia in a small city. More specifically, it tries to analyse the nature and extent of Mafioso infiltration into the legitimate private economy of the case-study using social network techniques. The following research question is tackled: in what kind economic sectors and companies does organised crime infiltrate?

The research will try to demonstrate that sectors with a high degree of centrality and comprising fewer firms (with consequently higher monopoly power) are the most vulnerable to this kind of criminal activity. It will also show that centrality is also the key criterion that makes a firm sensitive to infiltration, provided it belongs to a susceptible economic sector.

Such conclusions are reached through a four-step analysis. The first section will deal with the theoretical groundings. The research question will be contextualised by drawing on relevant economic, political and criminological literature. Organised crime will be conceptualised, as well as the concept of private legitimate economy. Based on previous research on the behavior of criminal groups, a model that explains why and how they infiltrate economy will be developed. The model will provide two predictions that will be tested in the case-study analysis. The first asserts that organised crime infiltrates sectors featuring high centrality in an economic network and high monopoly power, whereas the second advances that firms characterised by high centrality in those sectors are the most sensitive to criminal participation.

The second section will present the case study used to provide at least a partial answer to the research question. The empirical object of analysis will be the local private economy of a Sicilian city within a confined period, year 2002. The basic characteristics of the economy will be described, together with the Mafioso activity within it. Data on the criminal groups involved in legitimate businesses is provided by confidential material belonging to the Anti-Mafia Police. It will be discovered that one sector has been infiltrated by the local Mafia, and that five firms within it were connected to the criminal group.

---

[1] In this paper the terms organised crime, crime, Mafia, criminal group (and related adjectives) are used interchangeably and refer to the same subject of study.

The third section will test the two theoretical predictions using network analysis. A one-mode, non-directional network of those economic sectors active in the case-study will be constructed, where a tie between two nodes implicates the presence of economic transactions among them. An index that measures and weights degree centrality and monopoly power (proxied by the number of firms in the sector) of each node will be evaluated, and used to attempt to falsify the first prediction. In a similar fashion, a sub-network will be built, where nodes represent firms in the sector showing Mafioso activity, and edges are transaction among them. Degree centrality of each node will be computed and compared, such that the second prediction can be scrutinised by observing the results. An index of the degree of infiltration of the criminal group in the network will be also presented.

The fourth section will discuss the robustness of the results and the internal validity of the research. An instrument that adds uncertainty about the extent of penetration of the Mafia into the economy will be developed using statistical techniques, and some limitations of the research design will be emphasised. A conclusion will summarise the findings and suggest further research routes.

## 1. ORGANISED CRIME AND THE LEGITIMATE ECONOMY

The concept of organised crime has often escaped clarity in the scientific literature. This section aims at shedding light to the problem by introducing a precise definition of it, based on the latest juridical documents and criminological research. The other key concept of this research, that of private economy, is also delineated. Subsequently, a theory of why and how organised crime infiltrates is produced. Its assumptions – derived from bounded rationality and institutional economics – are presented, and its mechanisms explained.

### 1.1 *Defining organised crime*

The organised crime phenomenon is so ambiguous that more than 150 definitions of it exist (Van Lampe, 2011). The reasons do not involve only the already-mentioned methodological obstacles that prevent its proper empirical understanding, but also the different cultural and historical perspectives of scholars and policy-makers (Gurciullo, 2012). What organised crime means to a Dutch EU Commissioner might significantly differ from what a Bulgarian police officer thinks. Nonetheless, efforts towards an internationally recognised definition have been successfully performed by the European Union in 1992 (COE, 2002: 6), and later by the United Nations Convention against Organised Crime in Palermo (UN, 2000).

According to the latter document, this form of crime is best understood as

> "a structured group of three or more persons existing for a prolonged period of time and having the aim of committing serious crimes through concerted action by using intimidation, violence, corruption or other means in order to obtain, directly or indirectly, a financial or other material benefit" (UN, 2000).

However, it has been demonstrated that such a definition is inaccurate, because it encompasses forms of crime not commonly classified as organised crime. Examples are gangs' criminal activities (Pitts, 2007) and short-lasting forms of structured crime, such as a series of car thefts with several persons involved (Symeonidou-Kastanidou, 2007). Due to these methodological problems[2], this research adopts a more rigorous definition, developed by Gurciullo (2011a). It is shown in contrast to the UN one in table 1.

---

[2] There also exist critical legal and public policy arguments against the use of the UN definition. The reader is invited to read Finckenauer (2005) and Gurciullo (2011a) for a comprehensive discussion.

Table 1 – UN and the proposed definitions of organised crime

| | Properties | UN definition | Proposed definition |
|---|---|---|---|
| **Nature of the organisation** | Structured | ✓ | ✓ |
| | Three or more individuals engaged | ✓ | ✓ |
| | Prolonged period of activity | ✓ | ✓ |
| **Aims** | Serious crimes | ✓ | ✓ |
| | Material benefit and power | ✓ | ✓ |
| | Infiltration into the political system | ✗ | ✓ |
| **Methods** | Intimidation and violence | ✓ | ✓ |
| | Corruption | ✓ | ✓ |
| | Entrepreneurial planning | ✗ | ✓ |
| **Consequences** | Threat to stability of state | ✗ | ✓ |

Source: Gurciullo (2011a: 2)

In order not to mistake other forms of crime for Mafioso activities, three key connotations have been added. Other than the commission of serious crimes (that is, offences for which at least a three-year imprisonment is expected, such as drug and human trafficking [OSC, 2012]) and the pursuing of material benefits, organised crime aims at building connections with the political system of a country and influence it. This is exemplified by the Sicilian Mafia, which build a three-decade long symbiosis with main characters of the Italian Democratic Christian Party until the early 1990s (Lupo, 1996), or by the Taiwanese criminal group Heidao's political involvement. In 1999 half of the elected deputies were affiliated to the association (Chin, 2003: 15).

  A second key feature characterises the methods by which aims are reached. Violence and corruption are augmented by entrepreneurial-like forms of strategy. Division of labour and specialisation has become typical forms of performing tasks (Block, 1991: 11). The less internationally-known but most powerful Italian Mafioso group, 'Ndrangheta, succeeded in expanding its economic interests in Northern Italy thanks also to an efficient division between members who dealt with illicit affairs and those who laundered profits through legal investments (Varese, 2011: 31-64). Similar practices have been uncovered by Varese's (2012) content analysis of Russian investigative reports on the autochthonous criminal groups.

  A final property that distinguishes organised crime from other forms of crime is its inherent ability to pose as a threat to the stability of a state. Because of its aims and methods, heavy clashes with or distortion of the normal function of a government are inevitable. A dismal but evident example is found in the Mexican region of Tamaulipas, where the drug cartel Los Zetas has a virtually complete control of the territory and possesses a wide net of connection with politicians, public and police officers (Logan, 2009; Gurciullo, 2011b). The situation was no better in South Africa in the 1990s, where criminal syndicates kept engaging in a systematic corruption of officials that caused the dismantle of parts of the government administration (Minnaar, 1999).

1.2 *Defining legitimate private economy*

Conceptualising the notion of legitimate private economy is a much easier task, and can be done by describing the figure below.

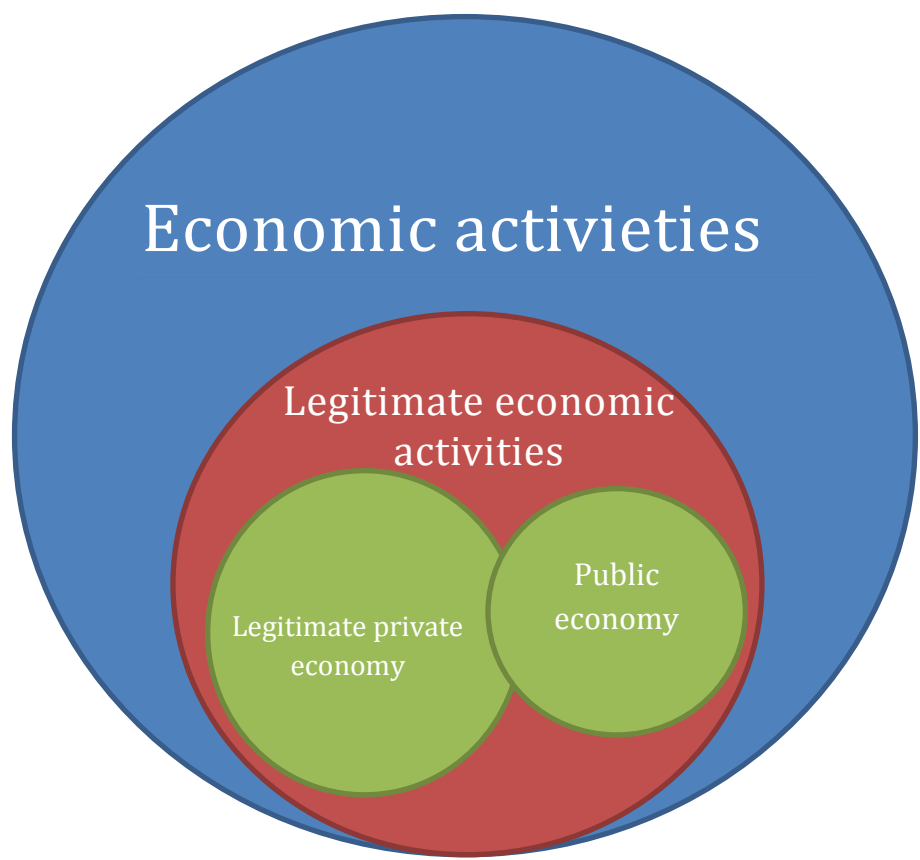
Figure 1 A taxonomy of the concept of economy

By economy it is meant a network of actors that deal with the management of human, natural and technological resources for the production, manufacture, trade and distribution of goods and services in a given territory (Mankiw and Taylor, 2006: 3), and is represented by the blue set. Indeed, this general definition incorporates all kinds of economic activities, from informal ones (not registered by any authority) to illicit ones (such as the buying and selling of stolen goods). The legitimate economy, therefore, is the purple subset, and is defined as the assemblage of economic activities that implicates deeds that are not punishable by the law of a state, and are formally recorded by an authority (ISTAT, 2006). Such activities might be performed by the public or by organisations owned by private individuals (or, in special cases, by a mixture of the two). If the latter is the case, a legitimate private economy is being observed.

A Mafioso association that infiltrates the legitimate private economy can be imagined as an actor working in the outskirts of the blue set injecting itself into the core of the larger green subset. The remaining parts of this section will illustrate why and how this is possible.

### 1.3 *How organised crime infiltrates: behavioural permissive causes*

A model of Mafioso economic infiltration logically requires at least three classes of elements or permissive causes. The first and most obvious one is the existence of private economic sectors in which organised crime can expand its influence. This is given for granted in the model and not further discussed.

The second class embraces the key presumptions about the behaviour and the decision-making strategy of organised crime. These properties explain the dynamic processes that eventually lead a criminal group to take possession of private companies. This research identifies three of them: a systematic tendency to maximise the aims of the criminal organisation, a bounded rational way of performing strategic decisions, and path-dependency. Their meaning and empirical validity is explained in the following paragraphs.

The last class comprises the set of possible events that trigger the criminal infiltration. It is theorised that the over-accumulations of capital and profits originated in the illicit markets may function as the catalysts for the phenomenon to occur. This component of the model is further explicated in sub-section 1.4. The final part, point 1.5, presents the bulk of the model and its predictions.

*Organised crime routinely strives to increase the chances of reaching its aims, or to expand the magnitude of the goals that are already attained.* Criminal groups are assumed to have an inherent tendency toward augmenting its ability to commit serious crimes (an end in itself) and to corrupt political systems so as to gain financial revenues (Becker, 1968) or power as authority and legitimisation (Sciarrone, 2006).

Instances that prove the empirical relevance of this basic property are ubiquitous. Santino (2007: 20) shows how the Sicilian Mafia directed its efforts towards international drug trafficking in the 1970s. At that time the mob had a virtually complete control of the economy and governmental apparatus of the Sicily, yet its leaders concentrated their efforts in seeking new frontiers for profit-making. They found an optimal opportunity in the then-emerging markets for heroin and cocaine (Lupo, 2008). Another interesting episode occurs in Japan. In 1990 its national government introduced a new measure to defy banking secrecy, as the Yakuza actively exploited the instrument in order to hide and launder profits coming from illicit activities. Contrarily to the expectations, the criminal group did not register any decrease in income in later years. On the opposite, it enjoyed increasingly growing amount of capital, and security agencies were not able to even lightly hamper the organisation (Friman, 1994).

*Organised crime is a bounded rational agent that deducts and tinkers*. In disaccord with the recent 'waves' of security and criminological research (Brown et al, 2000; Gottschalk, 2009: 77), this paper contends that the subject of study – as any forms of organisation, arguably (Simon, 1997) – cannot perform rational decision-making, that is, acting to maximise a goal after having perfectly analysed all possible alternatives of action (Gravelle and Rees, 2004: 6-7). Criminal organisations are not able to process all the available information in a short amount of time in order to make an 'optimal' choice. The reason is twofold. The first relates to the epistemic and aleatory uncertainty about the environment. In other words, information about contexts and factors that might influence a choice is so large, disperse or random that cannot be properly use to evaluate the probability of a certain event to happen (Huijbregts, 2011). The second limit is cognitive. Judgment biases or even the computational limits of human brains prevent rational decisions (Jones, 1999: 288-289).

Because of these constraints, organised crime is assumed to be only a bounded rational agent. Instead of directly maximising its goals, it arrives to them through a strategy based on a mixture of strategic decisions and trial-end-error attempts. Such a mode of behaviour has been best introduced by Beinhocker (2011: 409):

> The space of possibilities is too vast, the interactors themselves are too complex, their interactions with their environment are too complex.... [Organisations] are then left with no choice. They can use their powers of logic and deduction for as far as they will take them, but then at some point they need to try things, tinker and experiment, get feedback from the environment, and try again.

In a nutshell, a deductive-tinkering process lies at the core of Mafioso behaviour, and evidence for this argument is plenty. Gurciullo (2011c) extensively discusses how this phenomenon explains the arrival of Colombian drug cartels in Guinea-Bissau, turning the small and weak country into a narco-state. Because of heavy pressures from the United States and a growing demand for cocaine in Europe, South American drug lords decide to move their cocaine trade hub (formerly in the Caribbean) to West Africa. The choice was the result of years of acquiring information about possible country candidates, and experimentation with small-scale trafficking. In a similar vein, Italian Mafias performed a deductive-tinkering strategy to understand the best way of re-investing profits coming from the illegal markets in the 1970s. Eventually, they initiated a long-lasting partnership with financial trader Vito Roberto Palazzolo (Conti, 2009).

A final property that – it is assumed – characterises organised crime is *path-dependency*. The concept has been first introduced by economic historians and institutional economists (see Liebowitz and Margoliz, 2000; Garrouste and Iōannidēs, 2001), and is now utilised in several branches of the social sciences. This property implicates that if a given set of actions revealed to be successful in the pursuit of organised crime's aims, the organisation will most likely re-adopt such measures in the near future (Gurciullo, 2011b: 9). This assumption is logically connected with the previous two. If uncertainty and an imperfect mode of decision-making are the only possibilities, it is reasonable to think that an actor will endure with previous decisions (whose outcomes are known) as long as its effects do not change.

Evidence supporting the verisimilitude of path-dependency in organised crime is found in the degree of specialisation that some groups attain in order to survive and grow. The Turkish Mafia has purposely become the main exporter of heroin from Afghanistan and Pakistan to Europe, as this pattern of action resulted to be successful over the years in securing profits and power to the group (Bovenkerk and Yeşilgöz, 2004). The Italian Sacra Corona Unita – active in Apulia – specialised in the buying and selling of illegal weapons from the Balkan region (Vendola, 1996), and is now the main supplier of this illicit good to the other Italian Mafias and to Somali warlords (Gurciullo, 2012b).

1.4 *How organised crime infiltrates: triggering factors*

As already mentioned, a third class of elements is needed in the model in order to explain what leads criminal groups to undertake the penetration of an economy. Such elements would function as catalysts for a new pattern of action. In other words, they should be able to push the organisation away from path-dependency and bring it to utilise economic infiltration as a means to its goals. These triggering factors are the over-accumulation of capitals from the illicit markets.

Markets for illegal goods and services are an enormous source of profits because of two principal characteristics. Firstly, their supply is constrained by the high risks due to police prosecution, and the high costs incurred in organising an underground net of manufacture and transportation (Reuter and Kleiman, 1986). Secondly, the presence of few suppliers that retain their share of demand through violence allows monopolistic power (Fiorentini and Peltzmann, 1997). The results are extremely high prices with large profit margins.

Although valid empirical research on how much criminal organisations earn are difficult because of the very nature of the subject of study, some studies provide evidence for this phenomenon. A 2010 report from the Italian commerce organisation SOS Impresa (2010) estimated €70 billion profits per year for the Italian Mafias. Most of this money is believed to originate from the illicit drug market and racketeering. Humphreys (2011) attempted to evaluate Mexican organised crime's profit gains drawing on RAND reports, and concluded with an estimate of $10 billion.

1.5 *How organised crime infiltrates: final model and predictions*

The building blocks of the model have been discussed. How can they drive organised crime to infiltrate into the legitimate economy? This dynamic process goes as follows. First, a catalyst event occurs. The surge in illicit profits causes the over-presence of capital; thereby mafias are incentivised to use it to expand to yet-untouched markets (Gurciullo, 2011d). Second, this change in the state of affairs is deeply influenced by organised crime's aims. Will for further profits and the need to acquire more power and political influence eventually directs its deductive-tinkering decision process to see the legitimate private economy as an ideal candidate. However, it is argued that not all actors belonging to this type of economy are palatable: in fact, only specific kinds of sectors and companies are vulnerable.

A sector is a sub-network of an economy that is specialised in the production of a particular kind of good or service. In order to be considered for infiltration by organised crime, it must fulfill two criteria, logically deducted from the aims of the criminal organisation. First, the sector should be central to the economic network, that is, it must be connected with many other sectors through transactions. This enhances the chances for the Mafia to expand its influence to other areas of the economy. Second, the sector must be characterised by monopoly power. The higher the monopoly power within a sector, the higher is government control over it, due to anti-trust laws or similar regulation (Blair and Carruthers, 2010: 64-81). Thus, the chances for a criminal group to become acquainted with public institutions also increase.

Within the vulnerable sector, finally, organised crime will attempt to infiltrate companies that are more connected to others than the average. The motive is the same as in the case of sectors: higher centrality implies a higher probability of expanding one's own influence in the market.

Figure 2 A model of organised crime's economic infiltration

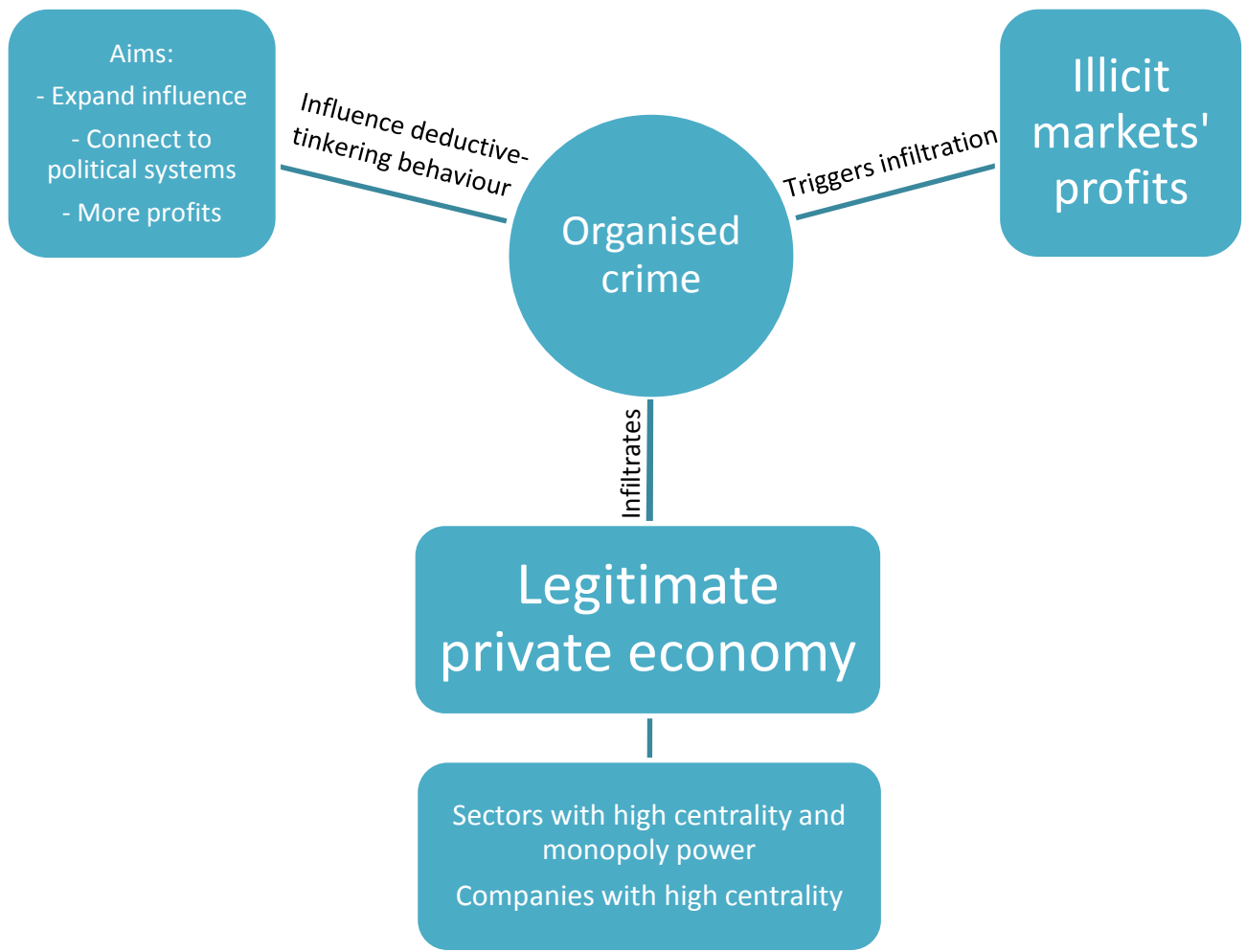

Figure 2 illustrates the model just proposed. In a nutshell, crime's huge illegal revenues spur a search for new investments, which eventually flow into the legitimate private economy, as it best helps fulfilling criminal organisations' aims. More specifically, Mafias penetrate sectors with high centrality and monopoly power, and chooses highly central firms in such sectors.

Although the academic literature on the theme is meager, parts of this theoretical model have already been subject to empirical test. Saviano (2006) and Veltri and Laudato (2009) provide detailed accounts of how money from black markets fomented legitimate entrepreneurial forms of organised crime. Calasanzio (2011) presents a series of case studies that show the Mafioso deductive-tinkering decision-taking approach leading to economic infiltration in Italy, while Jackson (2005) excellently proves the actual validity of this behavioural mode of action with regards to terrorist organisations. Despite the object of study differs, its findings can be easily transposed here.

What is lacking is a falsification test for the outcome of the model, that is, an analysis of whether infiltrated sectors and firms present the characteristics predicted above. This paper fills this gap by testing the hypotheses onto a case-study of a Sicilian city's economy, which is presented in the following section.

## 2. A CASE-STUDY: THE MAFIA IN A SICILIAN TOWN

An unprecedented access to confidential investigation documents authored by the Italian Anti-Mafia police allows the real-world observation of criminal economic infiltration. Firstly, the basic demographic characteristics of the case-study and its private economy are described. Subsequently, the economic Mafioso activities occurred in that area and discovered by authorities are discussed. The time of interest for the research is 2002, as investigation sources refer to activities happened in that period.

### 2.1 *The case-study: basic demographic and private economic characteristics*

The case analysed relates the economy of a small-size town locates in Southern Central Sicily, Porto Empedocle (a map with its location is shown in the Appendix). It hosts around 17,000 inhabitants (Comuni Italiani, 2012), although the population is experiencing small declines due to youth migration, just as many other small-size Italian communities (Di Zenzo, 2012).

Porto Empedocle's legitimate private economy can be operationalized and observed through the number of firms located and working within its territory. As table 3 shows, in 2002 1380 companies were active. Only 30.5% of these have 4 or more employees, suggesting that the local economy is mainly based on small-size, family owned businesses. This becomes clear by acknowledging that 0.05% of firms possessed more than 10 employees.

Table 2 Basic demographic and economic characteristics of the case-study

| Name | Porto Empedocle |
|---|---|
| **Population (2008)** | 17,261 |
| **Area** | 414,091 km$^2$ |
| **Number of firms** | 1380 |
| **Firms with more than 4 employees registered (%)** | 30.5 |
| **Firms with more than 10 employees registered (%)** | 0.05 |

Note: Data refers to year 2002 unless otherwise stated.
Sources: ISTAT (2009) and Camera di Commercio di Agrigento (2012a, 2012b, 2012c, 2012d, 2012e, 2012f).

Porto Empedocle's legitimate private economy is active in 29 sectors[3], as identified by Italy's National Chamber of Commerce (2012). Figure 2 decomposes the number of local firms according to the sector they work in. It is possible to notice that the great majority of companies is concentrated into sector 10 and 11, respectively that of retail and wholesale. This implicates that Porto Empedocle's inhabitants' main source of income is local commerce. Other sectors with a modest amount of firms are restaurant services (coded as 5), constructions (15), warehousing (22) and transportation of goods via land (29).

---
[3] Each sector is coded by a number. The reader is invited to read Section 3 for a complete account of the coding.

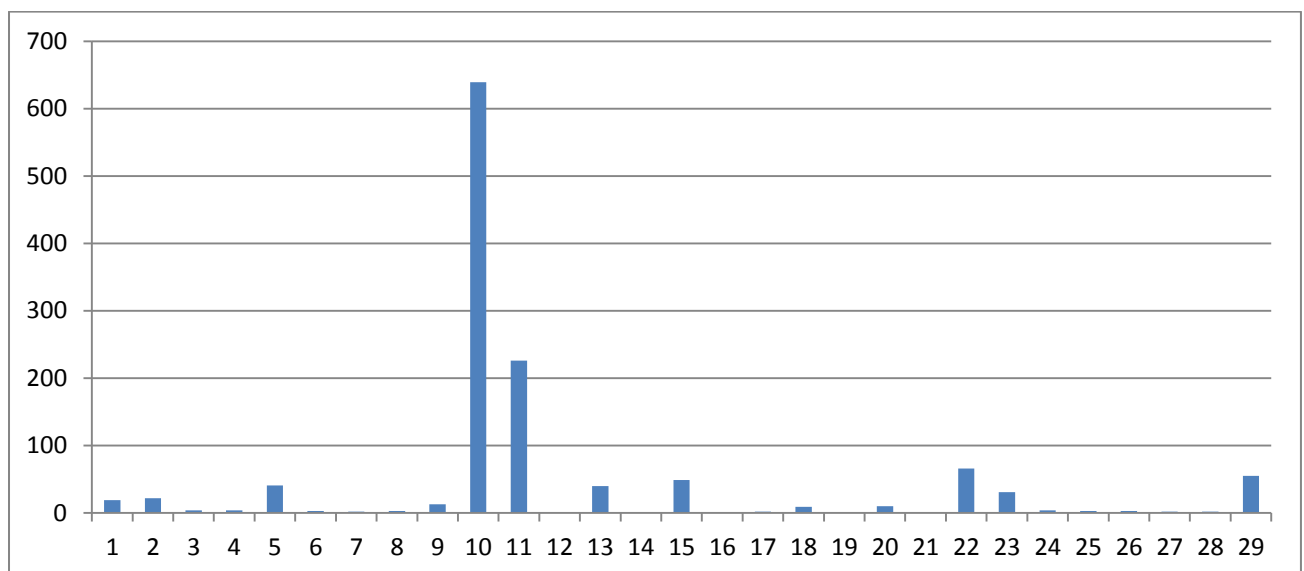

Figure 2 Number of firms according to economic sector

Source: Camera di Commercio di Agrigento (2012a, 2012b, 2012c, 2012d, 2012e, 2012f)

2.2 *The case-study: a narrative of the Mafioso infiltration*

Before proceeding to a quantitative analysis based on social network techniques, it is necessary to comprehend how organised crime infiltrated into the economy in a narrative mode. As already mentioned, all subsequent information originates from confidential documents and interviews with police personnel who preferred to remain anonymous. Names of persons and companies involved are not disclosed so as not to compromise further investigations.

Sicilian organised crime is better known as *Cosa Nostra* (literally, 'Our Thing'). Its structure is strictly hierarchical, with a *cupola* (the Italian word for 'dome') – composed by the most experienced and influential Mafioso members – taking large-scale strategic decisions, and *families* that retain authority in local territories. In-depth information about the institutional features of Cosa Nostra is found in Dickie (2004).

The Mafioso family controlling Porto Empedocle is the *Grassonellis*. According to investigative documents reporting facts occurred in 2002, they

> Operated in the construction sector [of Porto Empedocle] using intimidation and violence. These methods were aimed at constituting a monopoly in the production and supply of construction material in the territory of Porto Empedocle [...] and to force construction firms to buy the material they needed from it [the Mafioso firm] (translated from DDA, 2005a: 2).

In short, it is proven that the Mafia successfully acted to gain control of the construction sector of the economy of Porto Empedocle[4]. The infiltration started from the centre, that is, by becoming the actor that furnished all other firms with the material they utilised.

Confidential sources provide more details about this struggle for a violent monopoly. Allegedly, two entrepreneurs colluded with the Grassonelli family in order to render its construction firm Mafioso. DDA (2005b) supplies recorded conversations between the businessmen and a representative of Cosa Nostra. The suspects met a dozen times during the investigations, and conversed about the strategy to follow in order to acquire monopoly power in the market of interest. Typical themes were the listing and checking of firms that had to buy raw material from them, and the periodic payment of 'a living allowance' to the Grassonelli family by the colluded entrepreneurs (DDA, 2005b: 13-15).

---

[4] No definitive evidence of their penetration into other sectors is found, although still very plausible. The uncertainty about actual Mafioso economic activity is discussed in Section 4.

The Mafioso infiltration did not stop with the acquisition of firm supplying raw goods to other construction companies. There is clear evidence that Cosa Nostra constricted several firms to pay an extortion tax in order to stay in business.

In DDA (2005a: 68) it is cited a recorded conversation between the two Mafioso entrepreneurs. Worried about a potential competitor, they embark into the idea of killing its CEO. However, they decide not to perform such an act, as it might attract too much attention from the police, and because the measure should be used in 'more serious situations'. They finally agree to undermine the menacing firm using intimidation (i.e. the burning some of its vehicles) and obliging it to pay periodic sums of money in exchange for the safety of its workers.

A similar story is described by an arrested member of Cosa Nostra during an interrogatory. DDA (2005a: 10-12) admits that a firm specialised in the construction of sewage infrastructures bothered the Mafioso firm. After a series of small, intimidating attacks with Molotov bottles the victim company found itself negotiating a settlement with organised crime. Eventually, it agreed to pay an extortion tax to the Mafia and to avoid contracts that might interest the criminal group.

Wiretapping of Mafioso family members has allowed identifying two further Porto Empedocle's construction firms that had to withstand regular extortion taxes. These pressures were not recent, but most probably occurring since 1985 (DDA, 2005a: 44, 52). Indeed, this implies that the Mafioso influence into the legitimate economy is far from being a new phenomenon.

This section has shown how Mafioso infiltration occurred in a real case-study. In short, the strategy followed by the Grassonelli family consisted in finding willing businessmen to collude with, and to start with them a process aimed at gaining the most economic influence. While transforming a Mafioso firm into a monopoly managing raw construction materials, the group used violence so as to thwart competitors and making company pay sums of money. This permitted even greater influence in the sector. However, a chronicle is not enough to test the hypotheses advanced in section 2. The next section addresses this issue by performing a quantitative social network analysis of the legitimate private economy of Porto Empedocle.

## 3. SOCIAL NETWORK ANALYSIS OF THE CASE-STUDY

Social network analysis (SNA hereby) is the study of relationships among social entities through the use of graph theory and statistics (Wasserman and Faust, 2008: 3). Because of its nature it is the most adapt instrument to evaluate the model of this paper. This section goes as follows. The first part briefly overviews the usage of SNA in organised crime studies. The second formalises the model's predictions by turning them into hypotheses testable with SNA. The third one deals with the construction of the networks and how concepts are operationalized and data gathered. The final part tests the hypotheses and interprets the results.

### 3.1 Social network analysis in the organised crime literature

Research efforts with organised crime as the subject of analysis and SNA as the methodological tool are primarily characterised by a twofold nature. One field deals with the uncovering of organised crime's structure by identifying its nodes, types of relationships and overall network features. It can be both explorative (i.e. using an inductive process) and explanatory (that is, testing models of crime structure), and has greatly helped the understanding of this subtle phenomenon. McIllwain (1999) provided the theoretical justifications for this approach. An example of inductive use of SNA to understand criminal networks is Campana and Varese (2011), which shows how to unveil the internal structure of the Mafia by drawing data on wiretapping police reports. Easton and Karaivanov (2009), instead, endorse a deductive mode of action by testing the existence of optimal (that is, efficient and cost-effective) criminal networks. Other representative research is Morselli, Giguére and Petit (2007), Scaglione (2011) and Van der Hulst (2009).

The second branch departs from the findings of the former one and points at recognising the most critical parts of organised crime's structures. Its nature is more policy-oriented, directed towards the disruption of criminal links. Schwartz and Rouselle (2009) exemplify. Their paper exploits SNA to identify the central actors in a criminal network, and argues that they should be annihilated in order to mitigate organised crime threat. Another instance is Xia (2009), which proposes an adaptive inferring algorithm able to mine criminal networks' information and find critical nodes about which more investigation is needed. Further papers with similar methodological inclinations are Xu and Chen (2003, 2004).

It can be noticed that SNA applications in this area have only examined the *internal* structure of organised crime, and never its relationship with an *external* actor. This deficiency entails lack of information about how crime relates with other social agents. By looking at the connections between the legitimate private economy and the Mafia of small town, this paper represents a small step forward in the literature. It is hoped that further research would adopt and ameliorate the type of analysis explicated in the rest of this section.

### 3.2 *Formalisation of the hypotheses*

The theoretical model produced in Section 1 predicts that (1) organised crime infiltrates in those sectors with high centrality and monopoly power; and that (2) organised crime infiltrates in the most central firm(s) within such sectors. In order to formalise them, it is first necessary to translate them into statements referring to measures used in social network analysis.

Prediction (1) possesses two features, one referring to the position of the sector in the network, and the other to the very nature of sector. The first feature can be tested by using the concept of nodal degree – also known as degree centrality.

Let $\mathbb{N}$ be a network with $g$ actors and having a single, non-directional and binary type of relation. Each relation between two nodes $x_i$ and $x_j$ can therefore be either 0 or 1[5]:

$$x_{ij} = \begin{cases} 0 \\ 1 \end{cases}$$

Given this premise, the degree of a node can be defined as the sum of its actual non-zero relations over the total number of non-zero relations in that it can potentially have in $\mathbb{N}$[6]:

$$d(x_i) = \frac{\sum_{i=1}^{n} r_i}{\sum_{j=1}^{k} R_j} \quad (1)$$

Where $r_i$ refers to the values of set $r$, containing all the non-zero ties of $x_i$, while $R_j$ are elements of set $R$, which comprises all the non-null potential ties of the node in network $\mathbb{N}$[7].

A second measure – that of monopoly power – must be combined to the nodal degree in order to properly test the first prediction. Monopoly power in a market can be operationalised in several ways, the simplest of which is the number of companies working in it[8]. The smaller is the concentration of firms in a sector, the higher the likelihood that such a sector is typified by a monopolist form of market. Therefore, it must be created an index that weights sectors' centrality and concentration of firms.

As shown in figure 2, the number of firms may immensely change from one sector to another. This problem of extreme skewedness creates obstacles to the calculation of the index. It is solved through a linearisation of the variable. More specifically, each value can be ranked according to its position in the interquartile range of the series. Sectors with an extremely high concentration of firms are assigned lower numbers, and higher numbers are associated as the concentration diminishes. For the sake of consistency of the index, such a linearisation should be performed also on nodal degree (the higher the degree of a node, the higher is the number assigned to it), such that the index of centrality and concentration (ICC) is:

$$ICC = Rank\big(d(x_i)\big) + Rank\big(f(x_i)\big) \quad (2)$$

---

[5] The research adopts a sociometric notation. See Wasserman and Faust (2008: 77-83).
[6] The final step allows the normalisation of the nodal degree and it is not essential, although useful for a better representation of the results.
[7] It obviously follows that: $r \in R$
[8] There are more sophisticated indices of monopoly power (Elzinga and Millis, 2011; Rhoades, 1993; Saving, 1970), yet unavailability of data impedes their usage. A further discussion of the limits of this research is performed in the next section.

Where $d(x_i)$ and $f(x_i)$ are respectively the nodal degree of and number of firms in sector $x_i$. The beauty of equation (2) is that it allows a balanced weighting of the two factors involved in prediction (1)[9]. Thus, the latter can be rephrased in the following statement:

> $H_1$: Sectors penetrated by organised crime show a higher than average Index of Centrality and Concentration.

The reason for which $H_1$ does not endorse a more precise value of the ICC is that such a precision cannot be inferred by the theoretical model. Because of organised crime's deductive-tinkering approach, the penetrated sector may not be the optimal one, but only one that satisfices the criminal enterprise. A 'higher than average' proposition, thus, suffices and is logically consistent with and deducible from the theory.

The second prediction is easier to transpose, as it relates only to the centrality of Mafioso firms. It can therefore be stated:

> $H_2$: At least one of the firms experiencing Mafioso infiltration possesses the highest nodal degree in the sector's sub-network.

If both $H_1$ and $H_2$ are not rejected, a preliminary[10] confirmation of the empirical validity of the model is obtained.

### 3.3 *Networks construction and hypothesis testing*

Two networks have to be built. One shows the relationships among economic sectors in Porto Empedocle, whereas the other concentrates on the relation among firms in the construction sector – which is subject to Mafioso infiltration, according to the investigation files.

---

[9] The actual evaluation of the index is shown in part 3.3.
[10] Indeed, a case-study is not enough to definitively confirm or disconfirm the empirical validity of a model, although it represents a major step forward.

Figure 3 Network of economic sectors of in Porto Empedocle

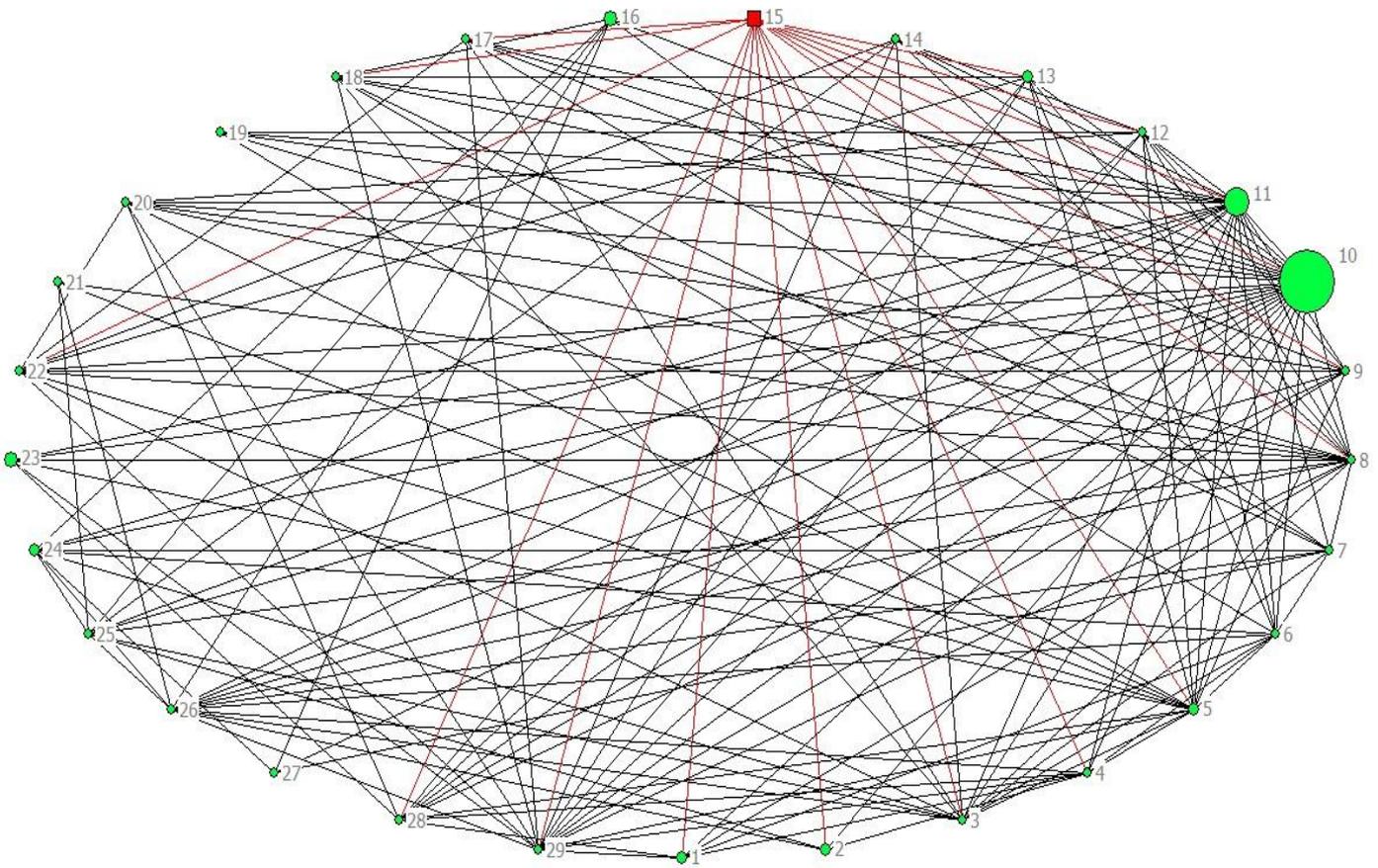

Both networks are easily fabricated with the UCINET software (Bargatti, Everett and Freeman, 2002). Figure 3 shows the first one. Each node is a sector that is active in the territory administered by the town. Ties are non-directional and binary in value, and exist if there has been at least one economic transaction between a firm in a given sector and a firm in another sector. The size of each node represents the number of firms in the sectors. As in figure 2, two large outliers (nodes 10 and 11) can be recognised. The shape and color indicates whether criminal infiltration has been detected by investigations. Only sector 15, the red squared one, shows this attribute. All the information dictated above is furnished by statistics recorded by the local Chamber of Commerce (Camera di Commercio di Agrigento, 2012a, 2012b, 2012c, 2012d, 2012e, 2012f) and refer to the year 2002.

Table 3 Summary Statistics of Sector Network

| ID | Description | Degree Centrality | Number of firms | Rank - Degree Centrality | Rank - Number of Firms | Index of Centrality and Concentration |
|---|---|---|---|---|---|---|
| 1 | Hotel and Hospitality | 0.25 | 19 | 4 | 4 | 8 |
| 2 | Personal Care and Services | 0.179 | 22 | 2 | 3 | 5 |
| 3 | Installation and Repair | 0.5 | 4 | 10 | 4 | 14 |
| 4 | Travel Services | 0.357 | 4 | 7 | 4 | 11 |
| 5 | Food and restaurant | 0.571 | 41 | 11 | 3 | 14 |
| 6 | Accounting and Finance | 0.429 | 3 | 9 | 5 | 14 |
| 7 | Media entertainment | 0.357 | 2 | 7 | 5 | 12 |
| 8 | Administration and Office support | 0.607 | 3 | 12 | 5 | 17 |
| 9 | Agriculture and Environment | 0.357 | 13 | 7 | 4 | 11 |
| 10 | Retail (except vehicles) | 0.893 | 639 | 14 | 1 | 15 |
| 11 | Wholesale (except vehicles and final consumption products) | 0.786 | 226 | 13 | 1 | 14 |
| 12 | Wholesale of final consumption product | 0.429 | 1 | 9 | 5 | 14 |
| 13 | Retail, Wholesale and Repair of Vehicles | 0.321 | 40 | 6 | 3 | 9 |
| 14 | Clothing Production | 0.286 | 1 | 5 | 5 | 10 |
| **15** | **Construction** | **0.571** | **49** | **11** | **3** | **14** |
| 16 | Electronic Manufacture | 0.250 | 1 | 4 | 5 | 9 |
| 17 | Chemical Products Manufacture | 0.286 | 2 | 5 | 5 | 10 |
| 18 | Metal Products Manufacture | 0.321 | 9 | 6 | 4 | 10 |
| 19 | Beverage Processing | 0.179 | 1 | 2 | 5 | 7 |
| 20 | Food Processing | 0.321 | 10 | 6 | 4 | 10 |
| 21 | Education | 0.143 | 1 | 1 | 5 | 6 |
| 22 | Warehousing | 0.393 | 66 | 8 | 2 | 10 |
| 23 | Fishing | 0.214 | 31 | 3 | 3 | 6 |
| 24 | Software Engineering | 0.321 | 4 | 6 | 4 | 10 |
| 25 | Computer Services | 0.357 | 3 | 7 | 5 | 12 |
| 26 | Postal Services | 0.500 | 3 | 10 | 5 | 15 |
| 27 | Telecommunications | 0.143 | 2 | 1 | 5 | 6 |
| 28 | Transportation via Sea | 0.393 | 2 | 8 | 5 | 13 |
| 29 | Transportation via Land | 0.607 | 55 | 12 | 3 | 15 |
| | **ICC arithmetic average** | | | | | **11.070*** |

*Number rounded to three decimal figures.

Degree centrality and the number of firms in each sector are illustrated in table 3, together with their respective rankings[11]. It can be observed that centrality indices are ranked on a scale from 1 to 13, where the highest value is held by the retail sector and the lowest by telecommunications. Firms' concentration is instead ranked on a smaller spectrum, spacing from 1 to 5. The reason for the difference lies in the lower diversity of values in the latter attribute. Both number series have been evaluated using a standard ranking algorithm found in Microsoft Excel software.

The last column exhibits the Index of Centrality and Concentration – the sum of the previous two ones. $H_1$ can therefore be subject to test. The node of interest is 15 and is highlighted in red. It presents an ICC value of 14, which is higher than the arithmetic average, 11.07. *$H_1$ is therefore not discarded* and the theoretical model of criminal economic infiltration is (at least partially) empirically valid.

Interestingly, the construction sector's ICC is placed third in the network, together with other sectors such as food and restaurant and wholesale. Sectors featuring higher ICCs are administration and office support, retail, postal services and transportation via land. The finding is extremely important, as it engenders at least two significant further statements to explore and that might enrich the theoretical model. The first one is that (1) criminal organisations are prone to infiltrate into all those sectors with an ICC equal or higher than the construction one; and (2) a further, unobserved attribute renders the construction sector more attractive to organised crime.

Unfortunately it is not possible to even argue for or against one of the two statements, as preliminary evidence provides empirical justification to both. Ecomafia (2009) is an extensive report about recent new Mafioso activities, and lists the infiltration into the retail and wholesale sectors and the new frontiers. In addition, Silj (1994: 360) contends that the reason why criminal organisations enter the construction market also lies in the possibility to collude with politicians, who usually regulate bids[12]. Further research is needed.

---

[11] Due to space constraints, the properties of this network cannot be extensively discussed, although this task may provide interesting points regarding the nature of local economies.

[12] Admittedly, this hypothesis could have been tested in this paper if non-fragmentary and good-quality data about the interactions between local politics and firms was available.

Figure 4 Network of firms in the construction sector

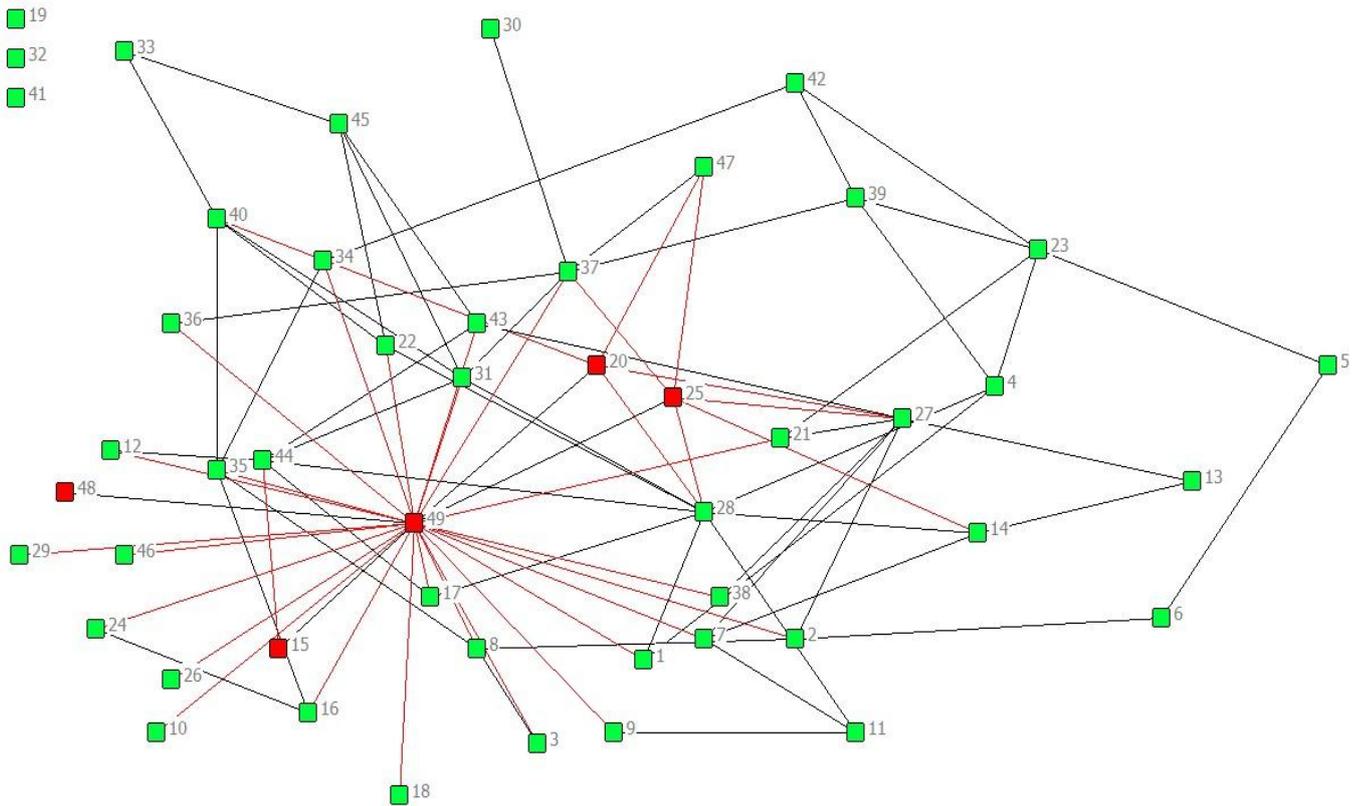

Figure 4 illustrates the second network of interest. Each of the 49 nodes represents a company in the construction sector of Porto Empedocle's private economy in 2002[13]. Analogously to the previous network, ties are non-directional and binary, and signify that at least one economic transaction has taken place between one firm and another in that year[14]. It might be legitimately asked why ties should exist in this network at all. If every node works in the same sector, they should be competitors and not experiencing such a form of exchange. Yet, this is not the case as construction firms may specialise in one of the many phases of the construction process, or in the construction of only certain kinds of structures. One company building a house might, for example, decide to bid the installation of windows and doors to other firms in the market.

Nodes in red are the ones subject to Mafioso infiltration according to the documents explored in sub-section 2.2. The estimation of their degree centralities allow the testing of $H_2$.

Table 4 Degree centrality of construction firms and Mafioso infiltration

| Nodes | Degree Centrality | Nodes | Degree Centrality |
|---|---|---|---|
| 49 | 0.583 | 34 | 0.063 |
| 28 | 0.208 | 42 | 0.063 |
| 27 | 0.167 | 47 | 0.063 |
| 37 | 0.146 | 3 | 0.042 |
| 25 | 0.125 | 5 | 0.042 |
| 31 | 0.125 | 6 | 0.042 |
| 44 | 0.125 | 9 | 0.042 |

---

[13] Names of companies are not disclosed in order to maintain confidentiality.
[14] Information about the number of firms and their transactions is given by Camera di Commercio di Agrigento (2012d).

*Table 4 (Continued)*

| | | | |
|---|---|---|---|
| 20 | 0.104 | 12 | 0.042 |
| 23 | 0.104 | 13 | 0.042 |
| 35 | 0.104 | 15 | 0.042 |
| 40 | 0.104 | 24 | 0.042 |
| 2 | 0.083 | 33 | 0.042 |
| 4 | 0.083 | 36 | 0.042 |
| 7 | 0.083 | 38 | 0.042 |
| 8 | 0.083 | 10 | 0.021 |
| 14 | 0.083 | 18 | 0.021 |
| 22 | 0.083 | 26 | 0.021 |
| 39 | 0.083 | 29 | 0.021 |
| 43 | 0.083 | 30 | 0.021 |
| 45 | 0.083 | 46 | 0.021 |
| 1 | 0.063 | 48 | 0.021 |
| 11 | 0.063 | 19 | 0.000 |
| 16 | 0.063 | 32 | 0.000 |
| 17 | 0.063 | 41 | 0.000 |
| 21 | 0.063 | | |
| **Mafioso Infiltration** | | | **0.233*** |

*Figure rounded to three decimal digits.
Note: Infiltrated companies are shown in red.

Table 4 presents degree centralities of construction firms sorted by magnitude. It can immediately be noticed that the most central company is a Mafioso one, 49. Thus, $H_2$ *is not rejected*. Company 49 furnishes concrete to all local construction firms using it. This explains its outstanding centrality index. Companies 20 and 25 show also relatively high connections with other nodes, while firms 15 and 48 are nearly isolate. The theoretical model does not provide definite explanations for this particular choice of firms, albeit it might be argued that the deductive-tinkering approach and contextual circumstances discussed in sub-section 2.2 eventually led Porto Empedocle's Mafia to disturb the activities of such businesses.

It is possible also to know the extent to which organised crime has infiltrated into the sector. This is easily done by evaluating the ratio between the sum of the ties of Mafioso firms and the sum of all relations in the network. This very simple index ranges from 0 to 1, where 0 implicates that no presence of criminal groups is known in the economic network, and 1 that the entire set of economic relations is influence by organised crime. At the bottom of table 4 it is estimated that the Mafia influences around 23% of economic relations in the network.

SNA of the case study has permitted to test the predictions advanced by the model of organised crime's infiltration developed earlier in this paper. Both hypotheses have not been refuted, indicating that the model possesses at least some empirical validity. Yet, several factors might undermine the internal validity of the study. These are explored in the next section, together with proposed research designs or modification that can overcome them.

## 4. THREATS TO INTERNAL VALIDITY

Internal validity relates to whether the inferences drawn from a research can be applied to the population of interest (King, Keohane and Verba, 1994). In this paper, this translates to whether the conclusions reached in the analysis of the previous section hold true in the case-study and, generally, to other instances of organised crime infiltration[15]. It is extremely important to clearly state any concern with the internal validity of research having an underground phenomenon such as organised crime as the object of study. Doing so allows future studies to bypass the encountered problems by carefully using the poor data available.

Two main issues that can impede internal validity are discussed below. These deal with uncertainty, and the operationalisation of the private economy and of economic relations.

4.1 *Uncertainty*

The impossibility of directly gather data and the very nature of covert criminal activities foment a quasi-inextricable veil of uncertainty regarding the inferences of the case-study analysis (Santino, 2006). Two sources of uncertainty are particularly critical.

The first appertains to the sector network. Investigation files provide extensive evidence regarding the Mafioso presence in the construction sector, yet it is very possible that organised crime penetrated into other markets. In DDA (2005b: 65) it is alleged that the Grassonelli family might also deal with the retail business, although no definitive proof has been gathered. In order to further understand the extent of uncertainty about Mafioso infiltration in Porto Empedocle's sectors, an Anti-Mafia police expert – who preferred not to disclose his identity – has been interviewed (Anonymous, 2012). He was asked to provide his opinion regarding the likelihood of other sectors being infiltrated by the local Mafia. The answer was that the event is extremely likely, because in several Southern Italian cities Anti-Mafia operations traced Cosa Nostra's investments in several types of market.

The implication is that inferences regarding the first hypothesis can be partly invalid: if other sectors are subject to Mafioso activities, and there is no information about their centralities and concentration of firms, it is not possible to ascertain whether $H_1$ has been refuted or not. This obstacle can be obviated by constructing an uncertainty interval showing minimum and maximum possible numbers of sectors penetrated. However, even minimal information about other sectors is not available or even produced; therefore such a measure of robustness cannot be developed.

The second source of uncertainty touches the construction sub-network and is extremely similar to the previous one. In short, it is likely that more firms have experienced organised crime's action, although the exact number is not known. In the interview (Anonymous, 2012), the anti-Mafia expert confirmed that investigations found proofs supporting the contention, but that they did not include it in the final report because such clues did not meet Italian Criminal Law standards for accusation[16].

Also this instance requires the construction of an uncertainty interval. Fortunately, it is possible to provide it thanks to the information held by the police expert.

---

[15] In this paper the population of study is comprised only of one observation, that of Porto Empedocle's Mafioso infiltration, while the population of interest is the set of other instances where a legitimate private economy has penetrated.

[16] It is worth noting that only a small part of investigation efforts manage to acquire such standards. Wiretapping or other single methods of proof-gathering are usually not enough. This implicates that many organised crime's activities are known but not punishable by law, and therefore not recorded in criminal reports presented to governmental authorities (Anonymous, 2012).

Figure 5 Mafioso infiltration's Gamma distribution

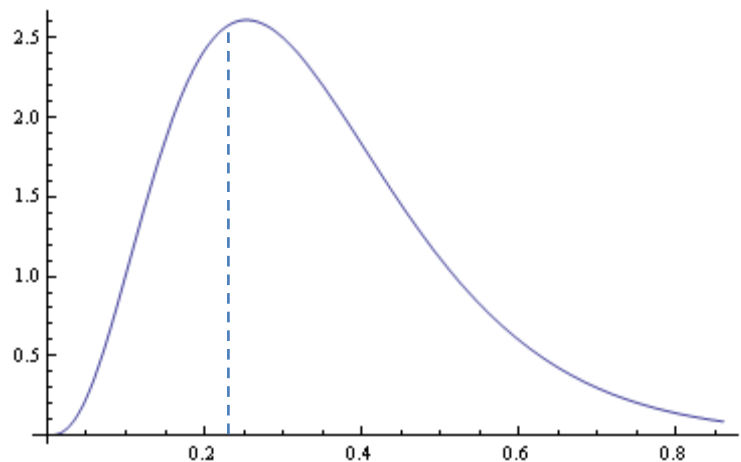

To construct a measure of uncertainty it must be assumed that the variable of interested is a random variable drawn from a known probability distribution[17]. Drawing on the police expert's declarations (Anonymous, 2012), it has been assumed that Mafioso infiltration in the construction market[18] (as evaluated in table 4) fits a Γ-distribution (Forbes et al., 2011: 109-113) of the following form:

$$M \sim \Gamma(3.915, 0.087)$$

Where M stands for the Mafioso infiltration index. The distribution has been calculated using software *Mathematica* and is drawn in figure 5. The choice of this particular type of function is justified by three key statements.

*It is extremely unlikely that firms found to be Mafioso are actually not Mafioso.* An extensive amount of evidence is taken over years of inspection, thus the chances that suspicious firms are innocent are extremely poor. This implies an extremely short tail at the left of the actual Mafioso infiltration index, whose value is 2.33.

*It is extremely likely that firms connected to the Mafia are more than the ones identified.* The assumption implicates that the distribution's mean is higher than the actual Mafioso infiltration index. In fact, trial-and-error attempts guided by the expert's information led the function to have a mean of 0.340.

*It gradually becomes more and more unlikely that Mafioso infiltration is higher than the mean.* The statement is justified by the fact that an overly penetrated market would leak more information about crime's activity and therefore be known. It implies a long distributional tail on the right of the mean.

---

[17] If more information were available, the distribution should be inferred rather than assumed.

[18] The index of Mafioso infiltration is a simple and easy-to-use measure of the infiltration of organised crime in the economy, as it is positively correlated to the number of firms penetrated. Another, more sophisticated candidate variable to construct an uncertainty interval on is the number of nodes in the network that are connected to crime. Yet, this would engender the use of random graphs techniques, which cannot be adequately implemented in this paper due to time and space constraints.

Table 5 Γ-distribution summary statistics

| | |
|---|---|
| **Shape** | 3.915 |
| **Scale** | 0.087 |
| **Mean** | 0.340 |
| **Variance** | 0.029 |
| **Standard Deviation** | 0.172 |
| **Skewness** | 1.011 |
| **Kurtosis** | 4.533 |

Note: figures rounded to the third decimal digit

Given the distribution is valid, it can finally be constructed an uncertainty interval. The interval would have a minimum value equal to the actual Mafioso infiltration, and a maximum amounting to the mean plus the standard deviation of the random variable[19], such that:

$$0.233 \leq M \leq 0.512$$

In other words, it can be said that although the actual infiltration is recognised to cover 23% of economic transactions, experts' opinion and the high likelihood that other firms are Mafioso could allow criminal influence to be as high as 51%, half of Porto Empedocles's construction market.

The addition of uncertainty does not directly compromise the non-rejection of $H_2$. Nevertheless, it allows inferences that develop further hypotheses and research questions. If more firms are victims of organised crime than actually known, what characteristics or patterns do they have in common? Do they follow the predictions stated in the theoretical model? Further research should deepen in this direction.

### 4.2 *Operationalisation of economic agents and economic relations*

Operationalisation is a process that identifies empirical variables that match or are directly correlated with a concept of interest. If it fails, erroneous data is used to make inferences, so that also inferences are wrong (King, Kehoane, and Verba, 1994; Adcock, 2001). The legitimate private economy has been operationalised with two indicators: economic sectors and actual firms. Yet, this sort of measurement might be incomplete: the legitimate private economy comprises not only producers but also consumers, thus, they should be added in the network reconstruction of the economy of the case-study. Consequences for $H_1$ and $H_2$ can be relevant: the centralities of sectors and firms might not be the same in a network where also consumers are contemplated. To test the robustness of the findings, information about consumers should be extracted, so that the networks built in the previous section would be augmented by one mode – becoming two-mode networks.

Economic relations among firms have been illustrated as ties showing whether at least one financial transaction occurred. It can be argued that the operationalisation is not accurate: one financial transaction over a year might not be enough to state that two actors are linked. This influences the number of ties the nodes have, and consequently the robustness of the findings regarding the case-study. To overcome the problem, networks can be modified in such a way as to have valued ties, which show the magnitude of transactions between a firm and another. However, this kind of data is not publicly available and should be procured by scrutinising all 2002 invoices of each firm in the case-study's economy.

---

[19] The range of the uncertainty interval depends on what range of possible values the researcher wants to be considered, and it is at his discretion. Here, the interval is set to cover 55% of possible values above the one actually measured.

## 5. CONCLUDING REMARKS

This paper has tried to uncover the patterns that characterise organised crime's infiltration into the legitimate economy though the social network analysis of a case-study. It has been found evidence that criminal associations tend to infiltrate into sectors having a high degree of centrality in the economy and a monopolist kind of market. Closer inspection has also demonstrated that firms with high centrality and located in vulnerable sectors are prone to be influenced by organised crime. These contentions have been approached after four phases of analysis.

Firstly, organised crime has been defined in a different and more precise way than usually done in the literature, so as to distinguish it from other forms of crime that are organised. Meanwhile, the concept of legitimate economy has been understood as the set of actors dealing with the production and consumption of good and services allowed by the law. A model of criminal economic infiltration has also been produced. Such a model assumes that organised crime is a path-dependent, bounded rational actor that routinely attempts to reach its goals. It shows that a catalyst series of events incites crime to enter the legitimate private economy, and releases two predictions to test empirically.

Secondly, the case-study, the economy of a Sicilian town named Porto Empedocle, has been introduced. Its principal economic features have been discussed, and the consultation of confidential investigation files authored by the Italian anti-Mafia police has helped confirming the nature of Mafioso infiltration. It has been established that a local Cosa Nostra family appropriated or influenced firms in the construction sector.

Thirdly, a social network analysis of the case-study has been performed. A network for the entire Porto Empedocle's legitimate private economy has been built (where nodes were economic sectors rather than firms), as well as one comprising only firms in the construction sector. An index weighting nodes' centrality and firms' concentration has been used to test the first hypothesis on the former network, while the simple calculation of degree centralities has been adopted to test the second hypothesis on the latter. Both postulates were not rejected, thereby augmenting the empirical validity of the theoretical model.

However, it has been considered opportune to warn the reader against possible threats to the validity of the inferences of the study. The fourth phase of the analysis has argued that a profound uncertainty about criminal economic activities in other sectors and the way economic actors and relations are operationalised can influence the final conclusions of the social network analysis. Unfortunately, it has been acknowledged that there is not enough information available (or resources to gather it) to make the study robust to such obstacles. Yet, thanks to experts' opinion it has been constructed a rudimental technique to estimate an uncertainty interval regarding the Mafioso infiltration in the construction sector.

This research has stepped into new terrain in the study of organised crime, and surely needs to be ameliorated in quality and augmented in scope. Three – out of many – future research routes are suggested below.

First of all, it is urgently needed to build a reliable model of uncertainty regarding organised crime economic activities. Because complete information about this phenomenon is unattainable (for reasons discussed throughout the paper), a measure of *how much it is not known* is essential. Future research could follow the technique used in section 4, and aims at constructing probability distributions of the variables of interest. Such a distribution could be estimated using Bayesian techniques taking into account experts' opinion and available information, together with theoretical knowledge (Cheeseman, 1983; Dempster, 2008a, 2008b). The same methods could be applied also in the generation semi-random networks, that is, networks where infiltrated sectors or firms and their relations change randomly within a pre-defined set of constraints (justified theoretically and/or empirically).

A second, essential, step forward is to add dynamics into the analysis. Only through the observation of criminal activities over time it is possible to provide more conclusive evidence for or against the theoretical model produced in the research. It would also allow to test if organised crime really follows a deductive-tinkering process in their quest for economic infiltration. Techniques substantially explained by Huisman and Snijder (2003), Kossinets and Watts (2006) and Snijider (2005) can guide towards this direction.

Finally, large-N studies are urgently necessitated. Models of Mafioso infiltration – and generally, the developing of theories and hypotheses – can properly be examined only through a diversification of the contexts in which variables are measured. In this way, any risk of selection bias is minimised. Further research should therefore design studies embracing several observations such as the one analysed in the case-study. The task is extremely difficult, and might be supplanted by substituting empirical samples with computerised ones – that is, with simulations of organised crime's infiltration. Gilbert and Doran (1994), Axelrod (2007) and Conte (2009) could provide the methodology to perform such path-breaking analyses.

# APPENDIX

Relevant questions asked during the interview with an anonymous anti-Mafia police expert (Anonymous 2012):

1. Why did the Grassonelli family decide to infiltrate the legitimate economy

2. Are other criminal organisations involved in the case study's economy?

3. Did the Mafia use the methods discussed in DDA (2005a, 2005b)?

4. What are the chances of Cosa Nostra being infiltrated in economic sectors other than the construction one? Could you give some examples?

5. Investigations identified five construction companies connected to the Mafia. How uncertain is this finding?

6. Is there any further information about Mafioso activity in other sectors?

7. Is there any further information about Mafioso activity in other construction companies?

8. Do you think that the uncertainty interval (section 4.1) reflects available information appropriately?